\documentclass[12pt,prd,tightenlines,nofootinbib,showpacs]{revtex4}
\newcommand{\be}{\begin{equation}}
\newcommand{\ee}{\end{equation}}
\newcommand{\bdis}{\begin{displaymath}}
\newcommand{\edis}{\end{displaymath}}
\newcommand{\bga}{\begin{equation}\begin{gathered}}
\newcommand{\ega}{\end{gathered}\end{equation}}
\usepackage{bm}
\usepackage{graphics}
\usepackage{rotating}
\usepackage{epsfig}
\usepackage{amsmath}
\usepackage{amsfonts}
\usepackage{amssymb}
\usepackage{makeidx}

\begin{document}

\title{Corrections of two-photon interactions in the fine and hyperfine 
structure of the P-energy levels of muonic hydrogen}

\author{A.~E.~Dorokhov\footnote{E-mail:~dorokhov@theor.jinr.ru}}
\affiliation{Joint Institute of Nuclear Research, BLTP,\\
141980, Moscow region, Dubna, Russia}
\author{\boxed{\rm{N.~I.~Kochelev}}}
\affiliation{Institute of Modern Physics of Chinese Academy of Sciences, 730000, Lanzhou, China}
\affiliation{Joint Institute of Nuclear Research, BLTP,\\
141980, Moscow region, Dubna, Russia}
\author{A.~P.~Martynenko\footnote{E-mail:~a.p.martynenko@samsu.ru}}
\affiliation{Samara University, 443086, Samara, Russia}
\author{F.~A.~Martynenko}
\affiliation{Samara University, 443086, Samara, Russia}
\author{A.~E.~ Radzhabov\footnote{E-mail:~aradzh@icc.ru}}
\affiliation{
Institute of Modern Physics, Chinese Academy of Sciences, Lanzhou 730000, China}
\affiliation{
Matrosov Institute for System Dynamics and Control Theory SB RAS, 664033, Irkutsk, Russia }

\begin{abstract}
In the framework of the quasipotential method in quantum electrodynamics
we calculate corrections to the nuclear 
structure proportional to $r_N^2$ from two-photon exchange amplitudes in the fine 
and hyperfine structure of P-states in muonic hydrogen, as well as the photon-photon interaction 
amplitudes, leading to the exchange of the axial vector meson. In constructing 
the quasipotential of the muon-nucleus interaction, we use the method of projection 
operators on states of two particles with a definite spin and total angular momentum. 
Analytical calculation of the matrix elements is performed and contributions to the 
fine and hyperfine structure of the $2P_{1/2} $ and $ 2P_{3/2}$ levels are obtained.
\end{abstract}


\maketitle

\section{Introduction}

The recent studies of the Lamb shift and hyperfine structure of the spectrum 
of muonic hydrogen, deuterium and helium have opened a new stage in precise investigation of the 
energy spectra of simplest atoms. Experiments performed by the CREMA collaboration (Charge 
Radius Experiments with Muonic Atoms) \cite{crema1,crema2,crema3,crema4} made 
it possible to carry out additional testing of the Standard Model, to obtain more 
accurate values of a number of fundamental parameters, and to raise the question 
of the presence of additional exotic interactions between particles in the bound 
state. Inclusion of other experimental groups in this area of research (see 
\cite{ma_2017,adamczak_2017,pohl_2017}) will allow not only to check the 
experimental results of the CREMA collaboration, but will also lead, as planned, 
to further increasing the accuracy of measurements of different intervals of 
fine and hyperfine structure.
The measurements already performed by the CREMA collaboration show that there is a significant 
difference between the values of such a fundamental parameter as the proton (deuteron) charge radius 
obtained in the study of the electron and muonic atoms \cite{crema3,crema4,CODATA,sick,carlson}.
It should be noted that recent study \cite{nnk} of the $(2S-4P)$ transition in the electronic 
hydrogen atom yielded a new value of the proton charge radius, $r_p = 0.8327(87)$~fm, 
which is quite consistent with the CREMA data. 
But new measurement of the transition frequency $(1S-3S)$ in hydrogen, carried out in 
Ref.~\cite{nez}, gave the value of the proton charge radius $r_p=0.877(13)$ fm, 
which completely agrees with the CODATA value.
As was often the case in long history of precision studies of the energy spectra 
of simplest atoms in quantum field theory, one of the ways to overcome the crisis 
situation involves a deeper theoretical analysis, testing previously 
calculated contributions that can play a large role in the case of muonic atoms. 
On this way, the problem of a more accurate theoretical construction of the particle 
interaction operator in quantum electrodynamics, the calculation of new corrections 
in the energy spectra of muonic atoms acquire a particularly important value 
\cite{apmplb,apm2014,apmplb2018,dorokhov2017,dorokhov2018,dorokhov2018a}.

In this paper, we investigate contributions of some two-photon interactions
in muonic hydrogen. The first part of two-photon exchange interactions is the corrections 
to the finite size of the nucleus in the fine and hyperfine structure of the P-energy levels.
If in the case of S-energy levels these two-photon
amplitudes have been studied in sufficient detail by different authors 
\cite{egs,gy,borie3,carlson2008,tomalak,tomalak2014,pineda1,pineda2}, then for P-energy levels such detailed calculations 
have not actually been carried out. Since the overlapping region of the muon wave 
function and the nucleus for P-states is significant, such contributions may be 
essential for a more accurate comparison of theory and experiment. This paper 
continues our series of studies of corrections to the structure of the nucleus 
without recoil \cite{apmplb,apm2014,apmplb2018} for the case of P-energy levels. 
It is well known that corrections to the structure and polarizability of the nucleus 
play an important role in the comparison of theory and experiment, since the experimental 
accuracy of measuring the atomic energy levels is constantly increasing. 
Periodically arising discrepancies between theory and experiment are usually attributed 
precisely to corrections for the structure of the nucleus, since there are always 
significant errors in their calculation.
The second part of the photon-photon interactions under consideration is related
with the appearance of an effective exchange of the axial vector meson
between the muon and proton which leads to 
shifts of P-levels of muonic hydrogen. In the case of S-states interactions of this type 
are significant \cite{dorokhov2017, dorokhov2018}.
The purpose of this part of the work is to understand how important the meson exchange 
processes can be to clarify the position of the P-energy levels.

\section{General formalism}

To investigate the fine and hyperfine structure of the energy spectrum of muonic 
hydrogen, we use a quasipotential method in quantum electrodynamics 
in which the bound state of a muon and a proton is described in the leading 
order in the fine-structure constant by the Schr\"odinger equation with 
the Coulomb potential \cite{apm2005,apm2015,pra2016}. The first part 
of important corrections in the energy spectrum of the S- and P-states 
is determined by the Breit Hamiltonian \cite{apm2005,apm2015,t4} 
(hereinafter abbreviated "fs" and "hfs" for the contribution to the 
energy spectrum of fine structure and hyperfine structure):
\begin{equation}
H_B=H_0+\Delta V_B^{fs}+\Delta V_B^{hfs},~~~H_0=\frac{{\bf p}^2}{2\mu}-\frac{Z\alpha}{r},
\label{eq:1}
\end{equation}
\begin{equation}
\label{eq:2}
\Delta V_B^{fs}=-\frac{{\bf p}^4}{8m_1^3}-\frac{{\bf p}^4}{8m_2^3}+\frac{\pi Z\alpha}{2}
\left(\frac{1}{m_1^2}+\frac{1}{m_2^2}\right)\delta({\bf r})-\frac{Z\alpha}{2m_1m_2r}
\left({\bf p}^2+\frac{{\bf r}({\bf r}{\bf p}){\bf p}}{r^2}\right)+
\end{equation}
\begin{displaymath}
+\frac{Z\alpha}{2m_1^2r^3}\left[1+\frac{2m_1}{m_2}+2a_\mu\left(1+\frac{m_1}{m_2}\right)\right]
({\bf L}{\bf s}_1),
\end{displaymath}
\begin{equation}
\label{eq:3}
\Delta V_B^{hfs}=\frac{8\pi\alpha\mu_p}{3m_1m_2}({\bf s}_1{\bf s}_2)\delta({\bf r})
-\frac{\alpha\mu_p(1+a_\mu)}{m_1m_2r^3}\left[({\bf s}_1{\bf s}_2)-3({\bf s}_1{\bf n})
({\bf s}_2{\bf n})\right]+
\end{equation}
\begin{displaymath}
\frac{\alpha\mu_p}{m_1m_2r^3}\left[1+\frac{m_1}{m_2}-\frac{m_1}{2m_2\mu_p}\right]({\bf L}{\bf s}_2)
\end{displaymath}
where $m_1$, $m_2$ are the masses of a muon and proton correspondingly, $\mu_p$ is the proton magnetic moment
in nuclear magnetons,
${\bf s}_1$ and ${\bf s}_2$ are the muon and proton spins. The contribution of interactions
(\ref{eq:1})-(\ref{eq:3}) to the energy spectrum of different muonic atoms was well investigated 
\cite{egs,borie3,kp1,uj,kkis,sgk2015,crema5,crema6}. The interaction operator 
(\ref{eq:3}) gives the basic contribution of order $\alpha^4$ to hyperfine structure of muonic
atom (the Fermi energy). The precise calculation of hyperfine structure, which is necessary for
a comparison with experimental data, requires the consideration of various corrections.

\begin{figure}[htbp]
\centering
\includegraphics{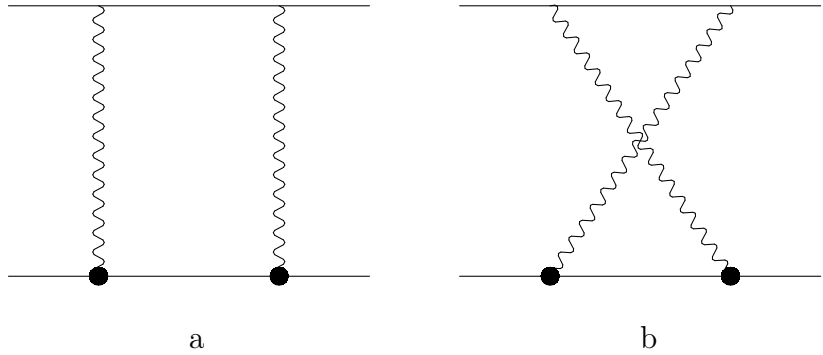}
\caption{Direct and crossed amplitudes of the two-photon exchange interaction. 
Thick dot in the diagram indicates the vertex operator of nucleus.}%
\label{fig1}%
\end{figure}

An infinite series of perturbation theory for the particle interaction operator 
contains contributions of different orders in $\alpha$ (the fine structure constant). 
One such contribution due to the two-photon exchange interaction is investigated 
in this paper. The amplitudes of this interaction are shown in Fig.~\ref{fig1}.

The two-photon exchange interaction amplitudes determine an important 
next to leading order contribution to the energy spectrum of muonic atom. 
The corrections of this interaction were studied in detail in the case of S-states both 
in the Lamb shift and in the hyperfine structure of the spectrum. The vertex of the 
electromagnetic interaction of a proton (nucleus) is determined by two electromagnetic 
form factors, information about which was obtained in the study of the scattering 
of leptons by protons (nuclei). Usually when studying this scattering the one-photon 
approximation is considered, in which the proton (nucleus) is in initial and 
final states on the mass shell. In the amplitudes in Figure~\ref{fig1} 
in each of the two vertices of the electromagnetic interaction there is a line 
of the nucleus, which denotes the virtual state of the proton (nucleus). 
Therefore, the use of free proton form factors at these vertices always caused 
criticism and led to theoretical errors, which in the case of muonic atoms 
are much larger than for electron atoms. 
At present, different methods are used to overcome these difficulties: 
the dispersion approach, effective field theories \cite{carlson2008,tomalak,tomalak2014,pineda1,pineda2}.
For P-states, the calculation 
of energy corrections was discussed to a much lesser degree than for the 
S-levels, so one of the goals of this paper is to fill this gap.

For the construction of the particle interaction operator
we use the tensor method \cite{apmplb,apm2014} based on the tensor representation 
of wave functions and projection operators for P-states of muonic hydrogen. 
It is convenient to work further in the momentum representation, in which the 
2P-state wave function of the muonic hydrogen atom has the form:
\begin{equation}
\label{eq:4}
\psi_{2P}({\bf p})=\left(\varepsilon\cdot n_p\right)R_{21}(p),
\end{equation}
where $\varepsilon_\delta(L)$ is the polarisation vector of orbital motion
with the normalization condition $\varepsilon^2=3/4\pi$, 
$n_p=(0,{\bf p}/p)$, $R_{21}(p)$ is the radial wave function in momentum representation.
Corrections to the energy levels are then determined in integral form:
\begin{equation}
\label{eq:5}
\Delta E(2^{2F+1}P_J)=\int\left(\varepsilon^\ast\cdot n_q\right)R_{21}(q)\frac{d{\bf q}}{(2\pi)^{3/2}}
\int\left(\varepsilon\cdot n_p\right)R_{21}(p)\frac{d{\bf p}}{(2\pi)^{3/2}} \Delta V({\bf p},{\bf q}).
\end{equation}
In the leading order the potential of fine or hyperfine structure 
$\Delta V({\bf p},{\bf q})$ is constructed by means of the amplitude of one-photon interaction
$T_{1\gamma}$ and is determined by formula \eqref{eq:3}. 
Starting with the two-photon interaction amplitudes, it is convenient to use the tensor method \cite{apmplb,apm2014} to construct 
the interaction potential. It consists in the fact that in the interaction amplitude 
of particles we consistently perform the projection on the muon states with 
the total angular momentum $ {\bf J} $ equal to 1/2 or 3/2, and then the projection 
of the muon-proton system to states with the total angular momentum $ {\bf F} = {\bf J} + {\bf s}_2 $. 
As a result of these actions, it appears that the contribution to the particle interaction operator 
for a given state is determined by the trace of the product of certain $\gamma$ factors, 
whose calculation and simplification can be performed for complex amplitudes using the 
Form \cite{form} package.
Let us clarify the realization of this calculation method using the amplitude example in Fig.~\ref{fig1}. 
Direct and crossed amplitudes have the following form:
\begin{equation}
\label{eq:6}
i{\cal M}_1=(Ze^2)^2\int\frac{d^4k}{(2\pi)^4}[\bar u_1(q_1)\gamma_\mu\frac{(\hat p_1+\hat k+m_1)}{(p_1+k)^2-m_1^2}
\gamma_\nu u(p_1)]\frac{g_{\mu\lambda}}{(k-p)^2}\frac{g_{\nu\sigma}}{(k-q)^2}\times
\end{equation}
\begin{displaymath}
[\bar v(p_2)\Gamma_\sigma\frac{(\hat k-\hat p_2+m_2)}{(k-p_2)^2-m_2^2}
\Gamma_\lambda v(q_2)],
\end{displaymath}
\begin{equation}
\label{eq:7}
i{\cal M}_2=(Ze^2)^2\int\frac{d^4k}{(2\pi)^4}[\bar u_1(q_1)\gamma_\mu\frac{(\hat p_1+\hat k+m_1)}{(p_1+k)^2-m_1^2}
\gamma_\nu u(p_1)]\frac{g_{\mu\lambda}}{(k-p)^2}\frac{g_{\nu\sigma}}{(k-q)^2}\times
\end{equation}
\begin{displaymath}
[\bar v(p_2)\Gamma_\lambda\frac{(-\hat k-\hat p_2+m_2)}{(k+p_2)^2-m_2^2}
\Gamma_\sigma v(q_2)],
\end{displaymath}
where $p_{1,2}=m_{1,2}v\pm p$ are four-momenta of a muon and proton in initial state,
$q_{1,2}=m_{1,2}v\pm q$ are four-momenta of a muon and proton in final state,
$v_\mu=(1,0,0,0)$. The proton vertex operators in direct and crossed amplitudes are equal
\begin{equation}
\label{eq:8}
\Gamma_\sigma=\gamma_\sigma F_1((k-p)^2)+\sigma_{\sigma\rho}\frac{(k-p)_\rho}{2m_2} F_2((k-p)^2), ~~~
\Gamma_\lambda=\gamma_\lambda F_1((k-q)^2)-\sigma_{\lambda\rho}\frac{(k-q)_\rho}{2m_2} F_2((k-q)^2),
\end{equation}
where $F_1(k^2)$ and $F_2(k^2)$ are the Dirac and Pauli electromagnetic form factors
of the proton.
When calculating the corrections to the structure of the nucleus without recoil (to the finite 
size of the nucleus) we can neglect in the proton propagator its kinetic energy:
\begin{equation}
\label{eq:9}
\frac{1}{(k-p_2)^2-m_2^2}\approx \frac{1}{-2m_2k^0+i0},~~~\frac{1}{(k+p_2-q-p)^2-m_2^2}\approx \frac{1}{2m_2k^0+i0}.
\end{equation}
In sum, these two expressions give the Dirac delta function $\delta(k^0)$, which allows us 
to perform an analytic integration with respect to $k^0$. Using the polarization vector 
$\varepsilon_\tau(L) $ of the orbital motion from \eqref{eq:4} and the muon Dirac bispinor $u(0) $ 
in the rest frame, we obtain for the $2P_{1/2}$ muon state at the first projection step the following 
expressions for numerators of direct and crossed amplitudes:
\begin{equation}
\label{eq:10}
N_{dir}=\frac{1}{3}[\bar\psi(0)(\gamma_\omega-v_\omega)\gamma_5(\hat q_1+m_1)\gamma_\mu(\hat p_1+\hat k+m_1)
\gamma_\nu(\hat p_1+m_1)\gamma_5(\gamma_\tau-v_\tau)\psi(0)\times
\end{equation}
\begin{displaymath}
\bar v(0)(\hat p_2-m_2)\Gamma_\sigma(\hat k-\hat p_2+m_2)\Gamma_\lambda(\hat q_2-m_2)v(0)],
\end{displaymath}
\begin{equation}
\label{eq:11}
N_{cr}=\frac{1}{3}[\bar\psi(0)(\gamma_\omega-v_\omega)\gamma_5(\hat q_1+m_1)\gamma_\mu(\hat p_1+\hat k+m_1)
\gamma_\nu(\hat p_1+m_1)\gamma_5(\gamma_\tau-v_\tau)\psi(0)\times
\end{equation}
\begin{displaymath}
\bar v(0)(\hat p_2-m_2)\Gamma_\lambda(\hat k-\hat p_2+m_2)\Gamma_\sigma(\hat q_2-m_2)v(0)],
\end{displaymath}
where the spinor $\psi(0)$ describes the muon in the rest frame with total momentum $j=1/2$.
Using the wave functions of the muon $\psi(0)$ and the proton $v(0)$, 
we introduce the projection operators on states with total angular momentum 
$F=1$ and $F=0$ in the second step. As a result, we obtain the following 
expressions in the numerator of the direct two-photon amplitude:
\begin{equation}
\label{eq:12}
N_{dir}^{(1)}=\frac{1}{24}Tr[\gamma_5(1+\gamma_0)(\gamma_\omega-v_\omega)\gamma_5(\hat q_1+m_1)
\gamma_\mu(\hat p_1+\hat k+m_1)\gamma_\nu(\hat p_1+m_1)
\end{equation}
\begin{displaymath}
\gamma_5(\gamma_\tau-v_\tau)(1+\gamma_0)\gamma_5
(\hat p_2-m_2)\Gamma_\sigma(\hat k-\hat p_2+m_2)\Gamma_\lambda(\hat q_2-m_2)]
\end{displaymath}
\begin{equation}
\label{eq:13}
N_{dir}^{(2)}=\frac{1}{72}Tr[\hat\varepsilon^\ast(1+\gamma_0)(\gamma_\omega-v_\omega)\gamma_5(\hat q_1+m_1)
\gamma_\mu(\hat p_1+\hat k+m_1)\gamma_\nu(\hat p_1+m_1)
\end{equation}
\begin{displaymath}
\gamma_5(\gamma_\tau-v_\tau)(1+\gamma_0)\hat\varepsilon
(\hat p_2-m_2)\Gamma_\sigma(\hat k-\hat p_2+m_2)\Gamma_\lambda(\hat q_2-m_2)],
\end{displaymath}
where the new polarization vector $\varepsilon_\mu$ in \eqref{eq:13} corresponds to the state with 
$F=1$. Similar expressions for \eqref{eq:12}-\eqref{eq:13} are obtained 
for crossed two-photon diagram. It should be emphasized that the advantage 
of this approach is that further transformations are performed within the Form 
\cite{form} package, which greatly simplifies the construction of the contribution 
to the particle interaction operator. As a result, the amplitude correction 
from \eqref{eq:6}-\eqref{eq:7} in the energy spectrum of $2^1P_{1/2}$, $2^3P_{1/2}$ states
can be represented in the following integral form:
\begin{equation}
\label{eq:14}
\Delta E(2^3P_{1/2})=\frac{(Z\alpha)^2}{96\pi^5 m_2}\int\frac{d\bf p}{(\bf k-\bf p)^2}R_{21}(p)
\int\frac{d\bf q}{(\bf k-\bf q)^2}R_{21}(q)\int\frac{d\bf k}{{\bf k}^2}\Bigl\{
F_1(({\bf k}-{\bf p})^2)F_2({\bf k}-{\bf q})^2)
\end{equation}
\begin{displaymath}
\times\bigl[\frac{4}{3pq}({\bf k}{\bf p})({\bf k}{\bf q})-\frac{4q}{3p}({\bf k}{\bf p})-
\frac{4p}{3q}({\bf k}{\bf q})+
\frac{2}{3}({\bf p}{\bf q})\bigl(\frac{p}{q}+\frac{q}{p}\bigr)\bigr]+
F_1(({\bf k}-{\bf p})^2)F_1(({\bf k}-{\bf q})^2)
\bigl[2{\bf k}^2\frac{({\bf p}{\bf q})}{pq}+
\end{displaymath}
\begin{displaymath}
\frac{4}{3pq}({\bf k}{\bf p})({\bf k}{\bf q})+
\frac{2q}{3p}({\bf k}{\bf p})(1+\frac{3m_2}{2m_1})+\frac{2p}{3q}({\bf k}{\bf q})(1+\frac{3m_2}{2m_1})+
\frac{5}{3}({\bf p}{\bf q})\bigl(\frac{p}{q}+\frac{q}{p}\bigr)+4m_1m_2\frac{({\bf p}{\bf q})}{pq}\bigr],
\end{displaymath}
\begin{equation}
\label{eq:15}
\Delta E(2^1P_{1/2})=\frac{(Z\alpha)^2}{32\pi^5 m_2}\int\frac{d\bf p}{(\bf k-\bf p)^2}R_{21}(p)
\int\frac{d\bf q}{(\bf k-\bf q)^2}R_{21}(q)\int\frac{d\bf k}{{\bf k}^2}\Bigl\{
F_1(({\bf k}-{\bf p})^2)F_2({\bf k}-{\bf q})^2)
\end{equation}
\begin{displaymath}
\times\bigl[\frac{4}{pq}({\bf k}{\bf p})({\bf k}{\bf q})-\frac{4q}{p}({\bf k}{\bf p})-
\frac{4p}{q}({\bf k}{\bf q})+
2({\bf p}{\bf q})\bigl(\frac{p}{q}+\frac{q}{p}\bigr)\bigr]+
F_1(({\bf k}-{\bf p})^2)F_1(({\bf k}-{\bf q})^2)
\bigl[-2{\bf k}^2\frac{({\bf p}{\bf q})}{pq}+
\end{displaymath}
\begin{displaymath}
\frac{4}{pq}({\bf k}{\bf p})({\bf k}{\bf q})-
\frac{q}{p}({\bf k}{\bf p})(6+\frac{m_2}{m_1})+\frac{p}{q}({\bf k}{\bf q})(6+\frac{m_2}{m_1})+
({\bf p}{\bf q})\bigl(\frac{p}{q}+\frac{q}{p}\bigr)-4m_1m_2\frac{({\bf p}{\bf q})}{pq}\bigr].
\end{displaymath}

In this work, we set the goal of investigating contributions to the finite size of the nucleus in the fine and hyperfine structure of the P-levels coming from two-photon exchanges.
To this end, we use an expansion of form  factors for small momenta of the form:
\begin{equation}
\label{eq:15a}
F_i({\bf k}^2)=F_i(0)+\frac{dF_i({\bf k}^2)}{d{\bf k}^2}\Bigr|_{\bf k^2=0}{\bf k}^2+
\frac{1}{2}\frac{d^2F_i({\bf k}^2)}{(d{\bf k}^2)^2}\Bigr|_{\bf k^2=0}{\bf k}^4+...
\end{equation}
In what follows, from this expansion we take into account only the terms $\sim r_E^2$, omitting the terms $\sim r_E^4$, and so on. After the expansion, our momentum integrals in \eqref{eq:14},
\eqref{eq:15} remain finite and there is no need to introduce any cutoff for them and divide the region of integration into a region of small and large momenta 
(In contrast at the calculation of the Zemach correction, such expansion leads to a divergent integral and it is necessary to divide the integration region into two parts of large and small momenta).
In the region of large momenta, the contribution of the integral is zero because of the angular integration. In the exact calculation of the resulting integrals (without expansion), the contribution of the region of large momenta is suppressed by additional factor $m_1^2/\Lambda^2$ and is of no interest for this study.

Expressions \eqref{eq:14}-\eqref{eq:15} are the result of the calculation of the amplitudes 
in the Form package with necessary accuracy.
To extract the contribution to the structure of the nucleus in the analytic form, 
we expand the form factors $F_{1,2}$ in a series, preserving the term proportional 
to the charge radius squared $r_E^2$. We take into account that the integral functions in 
\eqref{eq:14}-\eqref{eq:15} are symmetric with respect to the replacement 
${\bf p} \leftrightarrow {\bf q}$ and represent the product of form factors as follows:
\begin{equation}
\label{eq:16}
F_1(({\bf k}-{\bf p})^2)F_1(({\bf k}-{\bf q})^2)\approx  1+\frac{1}{3}r_E^2({\bf k}-{\bf p})^2,~
F_1(({\bf k}-{\bf p})^2)F_2(({\bf k}-{\bf q})^2)\approx  \kappa+\frac{\kappa}{3}r_E^2({\bf k}-{\bf p})^2,
\end{equation}
where $\kappa=F_2(0)$ is the proton anomalous magnetic moment, 
$r_E^2=-6\frac{dF_1(k^2)}{dk^2}\Bigr|_{k^2=0}
=-6\frac{dF_2(k^2)}{dk^2}\Bigr|_{k^2=0}/\kappa$.
Then the result of integration in \eqref{eq:14}-\eqref{eq:15} is defined by the following integral:
\begin{equation}
\label{eq:17}
I=\int\frac{d{\bf k}}{{\bf k}^2}\int pd{\bf p}R_{21}(p)\int \frac{d{\bf q}}{({\bf q}-{\bf k})^2}\frac{({\bf k}{\bf q})}{q}
R_{21}(q)=4\pi^5 W^4,~~~W=\mu Z\alpha,
\end{equation}
where $\mu=m_1m_2/(m_1+m_2)$.
Taking into account \eqref{eq:17}, we obtain from \eqref{eq:14}-\eqref{eq:15} analytical
expressions for the energy corrections:
\begin{equation}
\label{eq:18}
\Delta E(2^3P_{1/2})=E_F(2S)\frac{1}{24}(Z\alpha)^2m_1\mu r_E^2\left(-\frac{4\kappa}{1+\kappa}+\frac{3\frac{m_2}{m_1}+2}{1+\kappa}\right),
\end{equation}
\begin{equation}
\label{eq:19}
\Delta E(2^1P_{1/2})=E_F(2S)\frac{1}{8}(Z\alpha)^2m_1\mu r_E^2\left(-\frac{4\kappa}{1+\kappa}-\frac{\frac{m_2}{m_1}+6}{1+\kappa}\right),
\end{equation}
for convenience of presenting the result the Fermi energy for the 2S-level
is introduced: 
$E_F(2S)=\frac{1}{3}\frac{(1+\kappa)\mu^3\alpha^4}{m_1m_2}$.
Another scheme for adding moments involves first adding the particle spins to the total spin 
$S$, and then adding $S$ to the orbital angular momentum $L=1$. In this case, 
the transformation of the wave function of an atom has the form:
\begin{equation}
\label{eq:20}
\Psi_{JFF_z}=\sum_S(-1)^{F+S_1+L+S_2}\sqrt{2J+1}\sqrt{2S+1}\biggl\{
\begin{array}{ccc}
S_2 & S_1 & S \\
L & F & J
\end{array}
\biggl\}\Psi_{SFF_z}.
\end{equation}

The connection between the wave functions in a state with total angular momentum 
$F=1$ in two schemes for adding moments is determined as follows:
\begin{equation}
\label{eq:21}
\Psi_{\frac{1}{2}1F_z}=-\frac{1}{\sqrt{3}}\Psi_{01F_z}+\sqrt{\frac{2}{3}}\Psi_{11F_z},
\end{equation}
\begin{equation}
\label{eq:22}
\Psi_{\frac{3}{2}1F_z}=\sqrt{\frac{2}{3}}\Psi_{01F_z}+\frac{1}{\sqrt{3}}\Psi_{11F_z}.
\end{equation}
The contribution of states $\Psi_{01F_z}$, $\Psi_{11F_z}$ to the energy spectrum of P-levels has a form:
\begin{equation}
\label{eq:23}
\Delta E(2P,S=0,F=1)=\frac{3(Z\alpha)^2}{32\pi^5 m_2}\int\frac{d\bf p}{(\bf k-\bf p)^2}R_{21}(p)
\int\frac{d\bf q}{(\bf k-\bf q)^2}R_{21}(q)\int\frac{d\bf k}{{\bf k}^2}\times
\end{equation}
\begin{displaymath}
\Bigl\{F_1(({\bf k}-{\bf p})^2)F_2({\bf k}-{\bf q})^2)
\bigl[\frac{4}{3pq}({\bf p}{\bf q}){\bf k}^2-\frac{4}{3pq}({\bf k}{\bf p})({\bf p}{\bf q})-
\frac{4}{3pq}({\bf k}{\bf q})({\bf q}{\bf p})+
\frac{2}{3}({\bf p}{\bf q})\bigl(\frac{p}{q}+\frac{q}{p}\bigr)\bigr]+
\end{displaymath}
\begin{displaymath}
F_1(({\bf k}-{\bf p})^2)F_1(({\bf k}-{\bf q})^2)\bigl[2{\bf k}^2\frac{({\bf p}{\bf q})}{pq}+
\frac{1}{3pq}({\bf k}{\bf p})({\bf p}{\bf q})+
\frac{1}{3pq}({\bf k}{\bf q})({\bf p}{\bf q})(\frac{m_2}{m_1}-2)+
\end{displaymath}
\begin{displaymath}
\frac{1}{3pq}({\bf k}{\bf q})({\bf p}{\bf q})(\frac{m_2}{m_1}-2)+
({\bf p}{\bf q})\bigl(\frac{p}{q}+\frac{q}{p}\bigr)+4m_1m_2\frac{({\bf p}{\bf q})}{3pq}\bigr]\Bigr\}=
\end{displaymath}
\begin{displaymath}
=E_F(2S)\frac{1}{24}m_1\mu r_E^2(Z\alpha)^2\left(-\frac{4\kappa}{1+\kappa}+\frac{\frac{m_2}{m_1}-2}{1+\kappa}\right).
\end{displaymath}
\begin{equation}
\label{eq:24}
\Delta E(2P,S=1,F=1)=-\frac{3(Z\alpha)^2}{32\pi^5 m_2}\int\frac{d\bf p}{(\bf k-\bf p)^2}R_{21}(p)
\int\frac{d\bf q}{(\bf k-\bf q)^2}R_{21}(q)\int\frac{d\bf k}{{\bf k}^2}\times
\end{equation}
\begin{displaymath}
\Bigl\{F_1(({\bf k}-{\bf p})^2)F_2({\bf k}-{\bf q})^2)
\bigl[\frac{2}{3pq}({\bf p}{\bf q}){\bf k}^2-\frac{2}{3pq}({\bf k}{\bf p})({\bf k}{\bf q})\bigr]+
\end{displaymath}
\begin{displaymath}
F_1(({\bf k}-{\bf p})^2)F_1(({\bf k}-{\bf q})^2)\bigl[-\frac{2}{3pq}({\bf k}{\bf p})({\bf p}{\bf q})-
\frac{1}{6pq}({\bf k}{\bf p})({\bf p}{\bf q})(\frac{m_2}{m_1}+2)-
\end{displaymath}
\begin{displaymath}
\frac{1}{6pq}({\bf k}{\bf q})({\bf p}{\bf q})(\frac{m_2}{m_1}+2)-
\frac{q}{6p}({\bf k}{\bf p})(\frac{m_2}{m_1}+2)-\frac{p}{6q}({\bf k}{\bf q})(\frac{m_2}{m_1}+2)
-\frac{1}{3}({\bf p}{\bf q})\left(\frac{q}{p}+\frac{p}{q}\right)
-4m_1m_2\frac{({\bf p}{\bf q})}{3pq}\bigr]\Bigr\}=
\end{displaymath}
\begin{displaymath}
=E_F(2S)\frac{1}{12}m_1\mu r_E^2(Z\alpha)^2\frac{\frac{m_2}{m_1}+2}{1+\kappa},
\end{displaymath}
where when we add the spin and orbital angular momentum to the total moment $F=1$, 
the following tensor expression is used ($\epsilon^\beta$ is the polarization vector of the state with $F=1$):
\begin{equation}
\label{eq:25}
\varepsilon_{\rho\sigma}=\frac{i}{\sqrt{2}}\varepsilon_{\rho\sigma\alpha\beta}v^\alpha\epsilon^\beta.
\end{equation}
Using further \eqref{eq:21}-\eqref{eq:22}, we can obtain total contribution 
$\Delta E(2^3P_{1/2})$, $\Delta E(2^3P_{3/2}) $ from \eqref{eq:23}-\eqref{eq:24}. 
It is also necessary to take into account nondiagonal matrix element of the form
\begin{equation}
\label{eq:26}
\Delta E(S=0,S=1)=<\Psi_{01F_z}|\Delta V_{2\gamma}|\Psi_{11F_z}>=
\end{equation}
\begin{displaymath}
\frac{3(Z\alpha)^2}{32\sqrt{2}\pi^5 m_2}\int\frac{d\bf p}{(\bf k-\bf p)^2}R_{21}(p)
\int\frac{d\bf q}{(\bf k-\bf q)^2}R_{21}(q)\int\frac{d\bf k}{{\bf k}^2}
\Bigl\{F_1(({\bf k}-{\bf p})^2)F_2({\bf k}-{\bf q})^2)\times
\end{displaymath}
\begin{displaymath}
\bigl[-\frac{2}{3pq}({\bf k}{\bf p})({\bf p}{\bf q})+\frac{2q}{3p}({\bf k}{\bf p})+
\frac{2p}{3q}({\bf k}{\bf q})-\frac{2}{3pq}({\bf k}{\bf q})({\bf p}{\bf q})\bigr]+
\end{displaymath}
\begin{displaymath}
F_1(({\bf k}-{\bf p})^2)F_1(({\bf k}-{\bf q})^2)\bigl[\frac{1}{3pq}\frac{m_2}{m_1}({\bf k}{\bf p})({\bf p}{\bf q})+
\frac{1}{3pq}\frac{m_2}{m_1}({\bf k}{\bf q})({\bf p}{\bf q})-
\frac{q}{3p}({\bf k}{\bf p})\frac{m_2}{m_1}-\frac{p}{3q}({\bf k}{\bf q})\frac{m_2}{m_1}\bigr]=
\end{displaymath}
\begin{displaymath}
=E_F(2S)\frac{\sqrt{2}}{24}m_1\mu(Z\alpha)^2r_E^2\left(\frac{2\kappa}{1+\kappa}-\frac{{\frac{m_2}{m_1}}}{1+\kappa}\right).
\end{displaymath}
The contribution $\Delta E(2^3P_{1/2})$ has exactly the same form \eqref{eq:18}, 
as with another method of adding moments, and the contribution of this interaction 
to the level $2^3P_{3/2}$ shift is equal to
\begin{equation}
\label{eq:27}
\Delta E(2^3P_{3/2})=0.
\end{equation}

In the case of the state $2^5P_{3/2}$ with $F=2$, it is convenient to first add 
the spin moments of the particles, obtaining a state with spin $S=1$, which is 
described by the polarization vector $\varepsilon_\mu(S)$. Then, from the two wave 
functions $ \varepsilon_\mu(S)$ and $ \varepsilon_\omega(L)$, we construct 
the tensor $\varepsilon_{\mu \omega}$, describing the state with total 
angular momentum $F=2$. The further summation over the projections of total 
angular momentum is given by:
\begin{equation}
\label{eq:28}
\sum_\gamma\varepsilon^{\ast~(\gamma)}_{\mu\nu}\varepsilon^{(\gamma)}_{\lambda\omega}=
\frac{1}{2}\left({\cal P}_{\mu\lambda}{\cal P}_{\nu\omega}+
{\cal P}_{\mu\omega}{\cal P}_{\nu\lambda}\right)-\frac{1}{3}{\cal P}_{\mu\nu}{\cal P}_{\lambda\omega}.
\end{equation}

As a result the nuclear structure correction for the state $2^5P_{3/2}$ has the following
integral representation:
\begin{equation}
\label{eq:29}
\Delta E(2^5P_{3/2})=\frac{3\alpha^2}{32\pi^5 m_2}\int\frac{R_{21}(p)d{\bf p}}{({\bf p}-{\bf k})^2}
\int\frac{R_{21}(q)d{\bf q}}{({\bf q}-{\bf k})^2}\int \frac{d{\bf k}}{{\bf k}^2}\Bigl\{
F_1(({\bf k}-{\bf p})^2)F_1(({\bf k}-{\bf q})^2)\times
\end{equation}
\begin{displaymath}
\Bigl[-\frac{4}{3}{\bf k}^2\frac{{\bf p}{\bf q}}{pq}+\frac{2}{3}\frac{({\bf k}{\bf p})({\bf k}{\bf q})}{pq}-
\frac{5}{2}\frac{({\bf k}{\bf p})({\bf p}{\bf q})}{pq}\frac{m_2}{m_1}-9\frac{({\bf k}{\bf p})({\bf p}{\bf q})}{pq}+
{\bf k}{\bf p}\frac{q}{p}\left(\frac{5}{6}\frac{m_2}{m_1}+3\right)-
\end{displaymath}
\begin{displaymath}
\frac{({\bf k}{\bf q})({\bf p}{\bf q})}{pq}\left(\frac{5}{2}\frac{m_2}{m_1}+9\right)+
{\bf k}{\bf q}\frac{p}{q}\left(\frac{5}{6}\frac{m_2}{m_1}+3\right)-\frac{1}{3}({\bf p}{\bf q})
\left(\frac{q}{p}+\frac{p}{q}\right)-\frac{20}{3}m_1m_2\frac{({\bf p}{\bf q})}{pq}
\Bigr],
\end{displaymath}
\begin{displaymath}
F_1(({\bf k}-{\bf p})^2)F_2(({\bf k}-{\bf q})^2)\Bigl[
2{\bf k}^2\frac{{\bf p}{\bf q}}{pq}+\frac{2}{3}\frac{({\bf k}{\bf q})({\bf k}{\bf p})}{pq}-
4\frac{({\bf k}{\bf p})({\bf p}{\bf q})}{pq}-4\frac{({\bf k}{\bf q})({\bf p}{\bf q})}{pq}+
\end{displaymath}
\begin{displaymath}
\frac{4}{3}{\bf k}{\bf p}\frac{q}{p}+\frac{4}{3}{\bf k}{\bf q}\frac{p}{q}+\frac{4}{3}{\bf p}{\bf q}\left(\frac{p}{q}+
\frac{q}{p}\right)\Bigr]\Bigr\}.
\end{displaymath}
Using further the expansion of \eqref{eq:16} and extracting the correction to the nuclear structure ($\sim r_E^2$), 
we obtain a cancellation of all terms in \eqref{eq:29}.

The obtained expressions allow us to present a correction for the nuclear structure separately in the 
fine and hyperfine structure of the P-energy levels in the form:
\begin{equation}
\label{eq:30}
\Delta E(2P_{1/2})=\Delta E(2P_{1/2}(F=0))+\frac{3}{4}\Delta E^{hfs}(2P_{1/2}),
\end{equation}
\begin{displaymath}
\Delta E(2P_{3/2})=\Delta E(2P_{3/2}(F=1))+\frac{5}{8}\Delta E^{hfs}(2P_{3/2}),~
\Delta E^{fs}_{str}(2P_{3/2}-2P_{1/2})=-0.0014~\mu eV,
\end{displaymath}
\begin{equation}
\label{eq:31}
\Delta E^{hfs}_{str}(2P_{1/2})=0.2893~\mu eV,~~~
\Delta E^{hfs}_{str}(2P_{3/2})=0~\mu eV.
\end{equation}

It is useful to compare our results with the available calculations
of corrections to the finite size of the nucleus for P-levels in \cite{egs,friar,apm_2014}. 
For n = 2, using \cite{egs,friar,apm_2014}, we get $\Delta E(2P_{1/2})=(Z\alpha)^6r_E^2\mu^3/64$.
From our formulas \eqref{eq:18}, \eqref{eq:19}, \eqref{eq:23}, \eqref{eq:24}, 
\eqref{eq:26}, \eqref{eq:30} it follows that in the 
leading order in $m_1/m_2$ the analogous contribution is equal to
$\Delta E(2P_{1/2})=(Z\alpha)^6r_E^2\mu^3/48$
and differs from the previous one by the factor 4/3.
Nuclear size correction of order $m(Z\alpha)^6r_E^2$ was obtained in 
\cite{egs,friar,apm_2014} in $1\gamma$-interaction.
In this paper we have constructed the potentials of two-photon exchange, 
which effectively takes into account the contribution of the one-photon interaction, 
since the iterative term that we take into account is related with a point proton. 
Thus, the calculations of two-photon exchange amplitudes that have been carried out 
refine the previous result for P-levels of order $m(Z\alpha)^6r_E^2$.

When obtaining the numerical results in Table~\ref{tb1}, we use the value of the proton charge radius $r_E=0.84087$ fm. We note that the results (18), (19), (24) themselves are obtained in an analytical form, where the factor $r_E^2$ is extracted. Therefore, when comparing the theory and experiment and obtaining a more accurate value of $r_E$, these expressions can be used, forming a common coefficient before $r_E^2$ from all possible corrections.
Numerical values of the contributions \eqref{eq:28}-\eqref{eq:31} 
are presented in Table~\ref{tb1}. They refine the position of energy levels 
$ 2^1P_{1/2}$, $ 2^3P_{1/2}$, $2^3P_{3/2} $, $2^5P_{3/2} $ and may be 
needed for more exact comparison of theory and experiment. 
There is no contribution $\sim r_E^2$ to the energy of $2P_{3/2}$ state.
There is another type 
of two-photon interactions that contribute to the structure of P-states. It is 
considered in the next section.

\section{Contribution of axial vector mesons to muon-proton interaction}

\begin{figure}[th]
\centerline{
\includegraphics[scale=1.0]{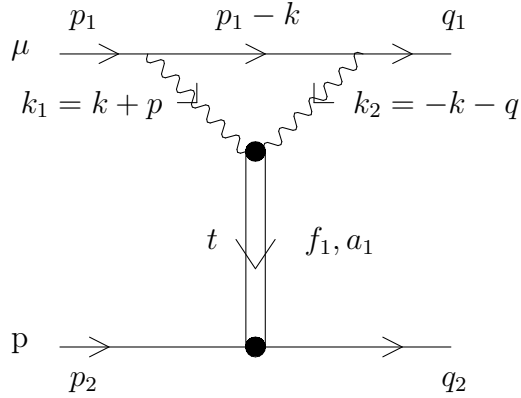}
}
\caption{Contribution of axial vector meson to the amplitude of muon-proton interaction.}%
\label{fig2}%
\end{figure}

Corrections in the fine and hyperfine structure of P-levels also arise as a result 
of two-photon interaction of another type, shown in Fig.~\ref{fig2}. In this case, 
the effect of two-photon interaction leads to the appearance of a meson exchange between 
the proton and the muon. The contribution of pseudoscalar mesons in the case of S-, 
P-states was calculated in \cite{pascalutsa,pang,kou,dorokhov2017}. The contribution 
of axial vector mesons to the hyperfine structure of the S-state spectrum, which was 
calculated in \cite{dorokhov2018}, is significant. Therefore, it is important to consider 
this contribution also to the $ 2P_{1/2} $, $2P_{3/2}$ level shifts.

An important role in the interaction amplitude in Fig.~\ref{fig2} is played by the vertex 
of the transition of two photons into an axial vector meson. For it, the following 
parameterization is valid \cite{cahn,pascalutsa2012}:
\begin{equation}
\label{eq:33}
T^{\mu\nu\alpha}=4\pi i\alpha\varepsilon_{\mu\nu\alpha\beta}\left(k_1^\beta k_2^2-k_2^\beta k_1^2\right)
F^{(0)}_{AV\gamma^\ast\gamma^\ast}(t^2,k_1^2,k_2^2),
\end{equation}
where $t=p-q$. If $k_1^2 = 0$, $k_2^2 =0$, then the vertex function \eqref{eq:33} vanishes. 
Introducing the projection operators to the state $2P_{1/2} $ as before in 
\eqref{eq:10}-\eqref{eq:11}, we represent the interaction amplitude of particles in this state in the form:
\begin{equation}
\label{eq:34}
{\cal M}(2^3P_{1/2})=\frac{\alpha^2g_{AVPP}}{512\pi^3m_1^2m_2^2}\int d^4kTr\Bigl\{
\hat\varepsilon(1+\gamma_0)(\gamma_\lambda-v_\lambda)\gamma_5(\hat q_1+m_1)
\gamma_\mu[m_1\hat v-\hat k+m_1]\gamma_\nu\times
\end{equation}
\begin{displaymath}
(\hat p_1+m_1)\gamma_5
(\gamma_\omega-v_\omega)(1+\gamma_0)\hat\varepsilon(\hat p_2-m_2)\gamma_\alpha
\gamma_5(\hat q_2-m_2) n^\lambda_q n^\omega_p F^{(0)}_{A\gamma^\ast\gamma^\ast}(t^2,k_1^2,k_2^2)\times
\end{displaymath}
\begin{displaymath}
\frac{\varepsilon_{\mu\nu\alpha\beta}}{k^2-2m_1k_0}\left[\frac{(k+p)_\beta}{(k+p)^2}+
\frac{(k+q)_\beta}{(k+q)^2}\right]\frac{1}{{\bf t}^2+M_A^2}
\Bigr\},
\end{displaymath}
where $g_{AVPP}$ is the coupling constant of the axial vector meson and proton.
To extract the leading order contribution in $\alpha$ we can make approximately the following replacements: 
$(k+p)^\beta/(k+p)^2\to (k^\beta+p^\beta)/k^2$, 
$(k+q)^\beta/(k+q)^2\to (k^\beta+q^\beta)/k^2$. Then the interaction amplitude gives
the contribution to the energy shift of $2P_{1/2}$ level, which can be presented in integral form:
\begin{equation}
\label{eq:35}
\Delta E(2^3P_{1/2})=\frac{\alpha^2g_{AVPP}}{4\pi^3}\int d{\bf p}R_{21}(p)\int d{\bf q}R_{21}(q)
\int\frac{dk}{k^2(k^2-2m_1k^0)}F^{(0)}_{AV\gamma^\ast\gamma^\ast}(0,k^2,k^2)\times
\end{equation}
\begin{displaymath}
\Bigl\{-\frac{8}{3}(kp)(kq)\frac{1}{pq}+\frac{4}{3}k_0^2({\bf p}{\bf q})\frac{1}{pq}
+\frac{2k_0}{m_1}pq-\frac{4}{3}pq\Bigr\}
\frac{1}{{\bf t}^2+M_A^2}.
\end{displaymath}
We left terms in brackets of expression \eqref{eq:35} that formally have a different order in $\alpha$, 
taking into account that $ p\sim \mu \alpha $, $ q \sim \mu \alpha $. The subsequent angular integration 
in the momentum space shows that their order is equalized and becomes equal to $\alpha^5$. 
All the momentum integrals in \eqref{eq:35} can be calculated analytically by using the vector dominance 
model for the function $ F^{(0)}_{AV \gamma^\ast\gamma^\ast}(0,k^2,k^2)$ \cite{vdm,L3C,L3Ca,aihara}:
\begin{equation}
\label{eq:36m}
 F^{(0)}_{AV \gamma^\ast\gamma^\ast}(0,k^2,k^2)=\frac{F^{(0)}_{AV \gamma^\ast\gamma^\ast}(0,0)}{\left(1+\frac{k^2}{\Lambda^2}\right)^2}=F^{(0)}_{AV \gamma^\ast\gamma^\ast}(0,0)F(k^2),
\end{equation}
\begin{equation}
\label{eq:36}
I_1=\int\frac{dk(k^2+2k_0^2)}{(k^4+4m_1^2k_0^2)}F^2(k^2)=
-\Lambda^2\frac{\pi^2}{2}\left[1+\left(\frac{5}{2}-2\ln 2+2\ln\frac{2m_1}{\Lambda}\right)
\frac{4m_1^2}{\Lambda^2}\right],
\end{equation}
\begin{equation}
\label{eq:37}
I_2=\int\frac{dk(3k_0^2-k^2)}{k^2(k^4+4m_1^2k_0^2)}F^2(k^2)=
4\pi^2\Bigl[-\frac{11}{48}-\frac{1}{8}\ln 2+\frac{1}{8}\ln\frac{2m_1}{\Lambda}+
\end{equation}
\begin{displaymath}
\frac{1}{96}\left(-31+24\ln 2-24\ln\frac{2m_1}{\Lambda}\right)\frac{4m_1^2}{\Lambda^2}
\Bigr],
\end{displaymath}
\begin{equation}
\label{eq:38}
J_1=\int d{\bf p}R_{21}(p)\int d{\bf q}R_{21}(q)\frac{{\bf p}{\bf q}}{pq}\frac{1}{({\bf p}-{\bf q})^2+M_A^2}=
\frac{2\pi^3 W^5}{(M_A+W)^4},~~~W=\mu Z\alpha,
\end{equation}
\begin{equation}
\label{eq:39}
J_2=\int pd{\bf p}R_{21}(p)\int q d{\bf q}R_{21}(q)\frac{1}{({\bf p}-{\bf q})^2+M_A^2}=
\frac{\pi^3 W^5M_A^2}{(M_A+W)^4}\left[3+\frac{4W}{M_A}+\frac{3W^2}{2M_A^2}
\right],
\end{equation}
where the values of parameter $\Lambda$ for three axial vector mesons $f_1(1285)$, $a_1(1260)$,
$f_1(1420)$ are equal 1.040 GeV, 1.040 GeV and 0.926 GeV correspondingly (see also our Table~I
from \cite{dorokhov2018}).
Here and below, we indicate the expansion of the exact result in powers of the parameter 
$m_1/\Lambda$ up to terms of the second order.
An important feature of the integral \eqref{eq:38} is that it is zero if 
$({\bf p}-{\bf q})^2$ is neglected in the denominator \eqref{eq:38}. 
This leads to an increase in the order of the contribution to $\alpha^5$. 
The total analytical result and numerical value of the contribution \eqref{eq:35} are equal to
\begin{equation}
\label{eq:40}
\Delta E(2^3P_{1/2})=\frac{\mu^5\alpha^7g_{AVPP}F^{(0)}(0,0)\pi^2}{M_A^2}\Biggl\{
-\frac{\Lambda^2}{M_A^2}\left[\frac{1}{6}+\frac{2m_1^2}{3\Lambda^2}\left(\frac{5}{2}+2\ln\frac{m_1}{\Lambda}\right)
\right]-
\end{equation}
\begin{displaymath}
-\frac{11}{12}+\frac{1}{2}\ln\frac{m_1}{\Lambda}+
\frac{1}{6}\left(-31-24\ln\frac{m_1}{\Lambda}\right)\frac{m_1^2}{\Lambda^2}
\Biggr\}=-0.0001~\mu eV.
\end{displaymath}
Numerical value of the contribution \eqref{eq:40} is obtained in the case 
of $f_1(1285)$ meson (see the values of corresponding parameters in 
\cite{dorokhov2018}). It is very small in comparison with the analogous contribution 
to the shifts of the S-levels, since the order of the contribution increases due to 
the additional factor $\alpha^2$. The contribution of another axial vector meson 
$a_1(1260)$ has approximately the same value, and the total contribution of all 
AV-mesons is shown in Table~\ref{tb1}. Similar results for the interaction 
amplitude and energy shift in the $2^1P_{1/2}$ state have the form:
\begin{equation}
\label{eq:40a}
{\cal M}(2^1P_{1/2})=\frac{\alpha^2 g_{AVPP}}{4\pi^3}\int\frac{dk F^{(0)}_{A\gamma^\ast\gamma^\ast}(k^2,k^2)}
{k^2(k^2-2m_1k_0)}\Bigl[-8\frac{(kp)(kq)}{pq}+4k_0^2\frac{{\bf p}{\bf q}}{pq}
+2pq\left(3\frac{k_0}{m_1}-2\right)\Bigr]\frac{1}{{\bf t}^2+M_A^2},
\end{equation}
\begin{equation}
\label{eq:40b}
\Delta E(2^1P_{1/2})=-\frac{\mu^5\alpha^7g_{AVPP}F^{(0)}(0,0)\pi^2}{2M_A^2}\Biggl\{
\frac{\Lambda^2}{M_A^2}\left[1+\frac{4m_1^2}{\Lambda^2}\left(\frac{5}{2}+\ln\frac{m_1}{\Lambda}\right)
\right]+
\end{equation}
\begin{displaymath}
+\frac{11}{2}-3\ln\frac{m_1}{\Lambda}+\frac{m_1^2}{\Lambda^2}
\left(31+24\ln\frac{m_1}{\Lambda}\right)
\Biggr\}.
\end{displaymath}
Numerical value of contribution \eqref{eq:40b} is presented in Table~\ref{tb1}.

To calculate the analogous contribution in the case of $2P_{3/2}$ state, 
we use the expansion \eqref{eq:22}, in which the projection on states 
$\Psi_{01F_z}$ and $\Psi_{11F_z}$ is also performed within the tensor method. 
As a result, the muon and proton interaction amplitudes for these states have the form:
\begin{equation}
\label{eq:41}
{\cal M}_{01F_z}=\frac{3\alpha^2 g_{AVPP}}{4\pi^3}\int\frac{dk F^{(0)}_{A\gamma^\ast\gamma^\ast}(k^2,k^2)}
{k^2(k^2-2m_1k_0)}\Bigl[\frac{4}{3}\frac{{\bf p}{\bf q}}{pq}(k_0^2+2k^2)+\frac{2}{3}\frac{({\bf p}{\bf q})^2}
{pq}\left(\frac{k_0}{m_1}-2\right)\Bigr]\frac{1}{{\bf t}^2+M_A^2},
\end{equation}
\begin{equation}
\label{eq:42}
{\cal M}_{11F_z}=\frac{3\alpha^2 g_{AVPP}}{8\pi^3}\int\frac{dk F^{(0)}_{A\gamma^\ast\gamma^\ast}(k^2,k^2)}
{k^2(k^2-2m_1k_0)}\Bigl[
\frac{2}{3}pq\frac{k_0}{m_1}-2k^2\frac{{\bf p}{\bf q}}{pq}+\frac{2}{3}\frac{k_0}{m_1}\frac{({\bf p}{\bf q})^2}{pq}
\Bigr]\frac{1}{{\bf t}^2+M_A^2}.
\end{equation}
The subsequent integration over the momenta $k$, $ {\bf p}$, $ {\bf q}$ in the calculation of the energy 
contributions can be performed analytically:
\begin{equation}
\label{eq:43}
I_3=\int\frac{dk k_0^2}{k^2(k^4+4m_1^2k_0^2)}F^2(k^2)=
\frac{\pi^2}{12}\Bigl[-7-6\ln\frac{m_1}{\Lambda}+\frac{2m_1^2}{\Lambda^2}\left(-23-24\ln\frac{m_1}{\Lambda}
\right)\Bigr],
\end{equation}
\begin{equation}
\label{eq:44}
I_3=\int\frac{dk k^2}{(k^4+4m_1^2k_0^2)}F^2(k^2)=
\frac{\pi^2}{3}\Lambda^2\Bigl[1+\frac{m_1^2}{\Lambda^2}\left(7+6\ln\frac{m_1}{\Lambda}
\right)\Bigr].
\end{equation}

\begin{table}[htbp]
\caption{Numerical values of corrections of two-photon interactions to the energy
spectrum of P-states in muonic hydrogen.}
\label{tb1}
\begin{center}
\centering
\begin{tabular}{|c|c|c|}   \hline
State   & Correction of $2\gamma$ exchange &  Correction of axial meson exchange \\
             & amplitudes, $\mu eV$ &  amplitudes, $\mu eV$ \\ \hline
$2^1P_{1/2}$    &  -0.2184    &  0.0005 \\  \hline
$2^3P_{1/2}$    &   0.0709   & -0.0002  \\  \hline
$2^3P_{3/2}$    &    0  &   -0.00005 \\  \hline
$2^5P_{3/2}$    &    0  &  0.00003 \\  \hline
\end{tabular}
\end{center}
\end{table}

When calculating the average values of axial vector meson exchange amplitudes 
by means of \eqref{eq:22}, it is necessary to calculate off-diagonal terms of the form
$\left\langle\Psi_{01F_z}|{\cal M}|\Psi_{11F_z}\right\rangle$. Denoting such a contribution
${\cal M}_{{01F_z},{11F_z}}$, we obtain the following expression for it:
\begin{equation}
\label{eq:45}
{\cal M}_{{01F_z},{11F_z}}
=\frac{3\alpha^2 g_{AVPP}}{4\sqrt{2}\pi^3}\int\frac{dk F^{(0)}_{A\gamma^\ast\gamma^\ast}(k^2,k^2)}
{k^2(k^2-2m_1k_0)}\Bigl[
-\frac{2}{3}pq\frac{k_0}{m_1}+\frac{2}{3}pq
+\frac{2}{3}\left(\frac{k_0}{m_1}-1\right)\frac{({\bf p}{\bf q})^2}{pq}
\Bigr]\frac{1}{{\bf t}^2+M_A^2}.
\end{equation}
Calculating then the contribution to the energy spectrum from \eqref{eq:45} as well as for the 
diagonal terms, we obtain:
\begin{equation}
\label{eq:46}
{\Delta E}_{{01F_z},{11F_z}}(2^3P_{3/2})=
\frac{\mu^5\alpha^7 g_{AVPP}\pi^2 F^{(0)}_{A\gamma^\ast\gamma^\ast}(0,0)}{3\sqrt{2}M_A^2}
\left(1-3\ln\frac{m_1}{\Lambda}+\frac{4m_1^2}{\Lambda^2}\right).
\end{equation}
Total numerical value for the level $2^3P_{3/2}$, which is determined by 
\eqref{eq:41}, \eqref{eq:42}, \eqref{eq:46}, is presented in
Table~\ref{tb1}. A similar contribution to the energy level $2^5P_{3/2}$ shift
can be obtained, if in the original amplitude of the muon-proton interaction 
we perform successive projection onto the state of particles with total spin 
$S=1$, then the state of total momentum $F=2$. In this case, the muon-proton 
state is described by the tensor $\varepsilon_{\mu\nu} $, which occurs when 
two polarization vectors $\varepsilon_{\mu}(L)$ and $\varepsilon_{\nu} (S)$ 
are added. Summation over the polarizations is carried out with the help of 
\eqref{eq:28}. After all the simplifications, the amplitude of particle interaction 
for $2^5P_{3/2}$ state is represented in the form:
\begin{equation}
\label{eq:48}
{\cal M}(2^5P_{3/2})=\frac{3\alpha^2 g_{AVPP}}{20\pi^3}\int\frac{dk F^{(0)}_{A\gamma^\ast\gamma^\ast}(k^2,k^2)}
{k^2(k^4-4m_1^2k_0^2)}\Bigl[
\frac{2}{3}pq(2k^2-3k_0^2)+\frac{2}{3}pq-
\end{equation}
\begin{displaymath}
\frac{1}{3}\frac{({\bf p}{\bf q})}{pq}k^2(13k^2+8k_0^2)+
2\frac{({\bf p}{\bf q})^2}{pq}(3k_0^2-2k^2)
\Bigr]\frac{1}{{\bf t}^2+M_A^2},
\end{displaymath}
and the corresponding contribution to the energy spectrum is included in Table~\ref{tb1}.

\section{Conclusion}

In this paper we study the contribution of some two-photon interactions to the muon 
and proton interaction potential and into the fine and hyperfine structure of the energy 
spectrum for P-states. In the framework of the quasipotential method in quantum electrodynamics, 
using the technique of projection operators on the states of two particles with a definite spin 
and total angular momentum, we construct the particle interaction operators \eqref{eq:14}, 
\eqref{eq:15}, \eqref{eq:34} and obtain analytical expressions for contributions to the 
fine and hyperfine structure of the P-energy levels. On their basis, numerical estimates 
of contributions are obtained, which we present in Table~\ref{tb1}. 
The method of projection operators allows us to represent the interaction amplitudes 
of particles in a covariant form and use computer methods for calculating Feynman 
amplitudes for a further construction of the particle interaction operator in a definite 
state, which is extremely important in the higher orders of perturbation theory.

A complete picture of the energy levels of muonic hydrogen with $n = 2$ 
was described in [3] in connection with the experiment of the CREMA collaboration.
We investigate corrections to the structure of the nucleus of order $\alpha^6$ 
proportional to $r_E^2$ in the case of P-levels. For the Lamb shift two-photon exchange 
amplitudes give a correction for the structure of the nucleus, which is expressed 
in terms of proton electromagnetic form factors (third Zemach). The same two-photon 
exchange amplitudes give another important contribution, which is known as a correction 
to the proton polarizability. In the case of the Lamb shift, the numerical values 
of these contributions to the structure and polarizability of the nucleus
are comparable in magnitude. In this connection, it can be expected that in the case 
of P-states the nuclear reactions of the electroproduction of $\pi$-, $\eta$-mesons and 
nucleon resonances can contribute comparable in magnitude to the considered here 
correction of order $\alpha^6$ to the structure of the nucleus.

The contributions of two-photon exchange amplitudes and amplitudes with the exchange 
of axial vector meson, studied by us, give essentially different numerical values of the 
P-level shifts. This is due to the fact that these two-photon interactions lead to corrections 
of different orders in the energy spectrum: two-photon exchange interactions contribute to a 
finite nuclear size of the order of $\alpha^6 $, and the exchange of the axial vector meson 
produces a contribution of order $\alpha^7$. In this connection, it can be said that the 
corrections to the structure of the nucleus from two-photon exchange amplitudes are more 
significant for a precise comparison with the experimental data because their numerical 
values are of the order of 0.0001 meV (see Table~\ref{tb1}). Recall that to explain 
the puzzle of the proton radius, a contribution of about 0.3 meV is needed. In our 
calculations of the contribution of axial vector mesons, we take into account the contributions 
of $ f_1(1285) $, $ a_1(1260)$, $f_1(1420) $ as in \cite{dorokhov2018}. 
Their total contribution is presented in Table~\ref{tb1}.
Theoretical uncertainties are not explicitly indicated in Table~\ref{tb1}. For the correction  
$\sim r_E^2$, the error is determined by the error in the value of the charge radius of the proton. 
If we take two values for the charge radius $r_E^{CODATA}=0.8775(51)$ fm and 
$r_E^{CREMA}=0.84087(39)$ fm, then the difference in the magnitude of this correction 
is about 9 percent. 
From the experimental data and model approximations for constructing the transition form factor and
the interaction potential of particles for axial vector exchange, we estimate the error in calculating 
the contribution of axial vector mesons approximately in 50 percent \cite{dorokhov2018}.

It should be noted that the axial vector meson exchange mechanism, which contributes to the 
hyperfine structure of the hydrogen atom spectrum, was discussed many years ago in 
\cite{kummer,nambu,drell}. The most important role for the numerical calculation of the contribution 
of light-by-light scattering is played by the form factor of the transition of two photons 
into an axial vector meson. We used for it, just like in the previous paper \cite{dorokhov2018}, 
the dipole parametrization with respect to the 4-momentum of each photon, obtained 
in the experiments of the collaboration L3 \cite{L3C,L3Ca,aihara}. 
The experimental data of L3 are limited by a rather small kinematic region, so new
measurements of transition form factors are extremely necessary. Such experiments can be 
performed by the BESIII and BELLEII collaborations.
The mechanism of interaction of a muon and a nucleus, which is related to effective meson exchange studied in this paper and also in previous publications \cite{dorokhov2017,dorokhov2018,dorokhov2018a}, 
is of great importance for the refinement of the position of the S- and P-energy levels of muonic atoms. Along with the exchange of pseudoscalar and axial vector mesons, it is of interest to consider the exchange of scalar and tensor light mesons, the number of which is significant. Our work in this direction is in progress.

\begin{acknowledgments}
The work is supported by Russian Science Foundation (grant No. RSF 18-12-00128) and
Russian Foundation for Basic Research (grant No. 18-32-00023) (F.A.M.).
\end{acknowledgments}


\begin{thebibliography}{99}
\bibitem{crema1}R.~Pohl, A.~Antognini, F.~Nez et al., Nature {\bf 466}, 213 (2010).
\bibitem{crema2}A.~Antognini et al., Science {\bf 339}, 417 (2013).
\bibitem{crema3}A.~Antognini et al., Ann. Phys. (NY) {\bf 331}, 127 (2013).
\bibitem{crema4}R.~Pohl, F.~Nez, L.~M.~P.~Fernandes et al., Science {\bf 353}, 669 (2016).
\bibitem{ma_2017} Y.~Ma et al., Int. J. Mod. Phys. Conf. Ser. {\bf 40}, 1660046 (2016).
\bibitem{adamczak_2017} A.~Adamczak et al. [FAMU Collaboration], JINST {\bf 11}, no. 05, P05007 (2016).
\bibitem{pohl_2017} R.~Pohl [CREMA Collaboration], J. Phys. Soc. Jap. {\bf 85}, 091003 (2016).
\bibitem{CODATA} P.~J.~Mohr, D.~ B.~Newell, and B.~ N.~Taylor (CODATA
Recommended Values of the Fundamental Physical Constants: 2014) Rev.
Mod. Phys. {\bf 88}, 035009 (2016).
\bibitem{sick}I.~Sick, Prog. Part. Nucl. Phys. {\bf 67}, 473 (2012).
\bibitem{carlson}C.~E.~Carlson, Prog. Part. Nucl. Phys. {\bf 82}, 59 (2015).
\bibitem{nnk}A.~Beyer et al., Science {\bf 358}, 79 (2017).
\bibitem{nez}H.~Fleurbaey et al. Phys. Rev. Lett. {\bf 120}, 183001 (2018).
\bibitem{apmplb}R.~N.~Faustov, A.~P. Martynenko, G.~A.~Martynenko and V.~V.~Sorokin, 
Phys. Lett. B {\bf 733}, 354 (2014).
\bibitem{apm2014}R.~N.~Faustov, A.~P. Martynenko, G.~A.~Martynenko and V.~V.~Sorokin, 
Phys. Rev. A {\bf 90}, 012520 (2014).
\bibitem{apmplb2018}R.N.~Faustov, A.P.~Martynenko, F.A.~Martynenko, and V.V.~Sorokin,
Phys. Lett. B {\bf 775}, 79 (2017).
\bibitem{dorokhov2017}A.~E.~Dorokhov, N.~I.~Kochelev, A.~P.~Martynenko, F.~A.~Martynenko, and R.~N.~Faustov,
Phys. Particles and Nuclei Lett., {\bf 14}, No.6, 857 (2017); arXiv:1704.07702 [hep-ph].
\bibitem{dorokhov2018}A.~E.~Dorokhov, N.~I.~Kochelev, A.~P.~Martynenko, F.~A.~Martynenko, and A.~E.~Radzhabov,
Phys. Lett. B {\bf 776}, 105 (2018).
\bibitem{dorokhov2018a}A.~E.~Dorokhov, N.~I.~Kochelev, A.~P.~Martynenko, F.~A.~Martynenko, A.~E.~Radzhabov,
and R.~N.~Faustov, J. Phys.: Conf. Series {\bf 938}, 012042 (2017).
\bibitem{egs}M.~I.~Eides, H.~Grotch and V.~A.~Shelyuto,
Theory of Light Hydrogenic Bound States, Springer Tracts in Modern Physics,
V. 222 (Springer, Berlin, Heidelbeg, New York, 2007).
\bibitem{gy}H.~Grotch and D.R.~Yennie, Rev. Mod. Phys. {\bf 41}, 350 (1969).
\bibitem{borie3}E.~Borie, Ann. Phys. {\bf 327}, 733 (2012).
\bibitem{carlson2008}C.E.~Carlson and M.~Vanderhaeghen, Annu. Rev. Nucl. Part. Sci.
{\bf 57}, 171 (2007).
\bibitem{tomalak}O.~Tomalak and M.~Vanderhaeghen, Eur. Phys. J. A {\bf 51}, 24 (2015).
\bibitem{tomalak2014}O.~Tomalak and M.~Vanderhaeghen, Phys. Rev. D {\bf 90}, 013006 (2014).
\bibitem{pineda1}C.~Peset and A.~Pineda, Nucl. Phys. B {\bf 887}, 69 (2014).
\bibitem{pineda2}C.~Peset and A.~Pineda, JHEP {\bf 04} (2017) 060.
\bibitem{apm2005}A.~P.~Martynenko and R.~N.~Faustov, J. Exp. Theor. Phys. {\bf 98}, 39 (2004).
\bibitem{apm2015}R.~N.~Faustov, A.~P.~Martynenko, G.~A.~Martynenko, and V.V. Sorokin, 
Phys. Rev. A {\bf 92}, 052512 (2015).
\bibitem{pra2016}A.~A.~Krutov, A.~P.~Martynenko, F.~A.~Martynenko, and O.~S.~Sukhorukova,
Phys. Rev. A {\bf 94}, 062505 (2016).
\bibitem{t4}V.~B.~Berestetskii, E.~M.~Lifshits and L.~P.~Pitaevskii, {\it Quantum
Electrodynamics}, (Nauka, Moscow, 1980).
\bibitem{kp1}K.~Pachucki, Phys. Rev. A {\bf 53}, 2092 (1996).
\bibitem{uj}U.~D.~Jentschura, Ann. Phys. {\bf 326}, 500 (2011).
\bibitem{kkis}S.~G.~Karshenboim, V.~G.~Ivanov, E.~Yu.~Korzinin, and V.~A.~Shelyuto,
Phys. Rev. A {\bf 81}, 060501 (2010).
\bibitem{sgk2015}S.~G. Karshenboim, E.~Yu.~Korzinin, V.~A.~Shelyuto, and V.~G.~Ivanov,
J. Phys. Chem. Ref. Data {\bf 44}, 031202 (2015).
\bibitem{crema5}B.~Franke et al., Eur. Phys. J. D {\bf 71}, 341 (2017).
\bibitem{crema6}J.J.~Krauth et al., Annals Phys. {\bf 366}, 168 (2016).
\bibitem{form}J.~A.~M.~Vermaseren, FORM, arXiv:math-ph/0010025.
\bibitem{friar}J.L.~Friar, Ann. Phys. {\bf 122}, 151 (1979).
\bibitem{apm_2014}A.P.~Martynenko, A.A.~Krutov and R.N.~Shamsutdinov, 
Phys. Atom. Nucl. {\bf 77}, 786 (2014).
\bibitem{pascalutsa}F.~Hagelstein and V.~Pascalutsa, PoS CD15 077 (2016).
\bibitem{pang}H.~Q.~Zhou and H.~R.~Pang, Phys. Rev. A {\bf 92}, 032512 (2015).
\bibitem{kou}N.~T.~Huong, E.~Kou and B.~Moussallam, Phys. Rev. D {\bf 93}, 114005 (2016).
\bibitem{cahn}R.~N.~Cahn, Phys. Rev. D {\bf 35}, 3342 (1987).
\bibitem{pascalutsa2012}V.~Pascalutsa, V.~Pauk and M.~Vanderhaeghen, Phys. Rev. D {\bf 85}, 116001 (2012).
\bibitem{vdm}G.~A.~Schuler, F.~A.~Berends and R.~van Gulik, Nucl.\ Phys. B {\bf 523}, 423
(1998).
\bibitem{L3C}P.~Achard et al. [L3 Collaboration], Phys. Lett. B {\bf 526}, 269 (2002).
\bibitem{L3Ca}P.~Achard et al. [L3 Collaboration], JHEP 03, 018 (2007)
\bibitem{aihara}H.~Aihara et al., Phys. Rev. D {\bf 38}, 1 (1988).
\bibitem{kummer}D.~F.~Flamm and W.~Kummer, Nuovo Cim. {\bf 28}, 1579 (1963).
\bibitem{nambu}S.~Fenster, R.~Koberle and Y.~Nambu, Phys. Lett. {\bf 19},
513 (1965).
\bibitem{drell}S.D.~Drell and J.D.~Sullivan, Phys. Lett. {\bf 19}, 516 (1965).
\end{thebibliography}
\end{document}